\documentclass[journal]{IEEEtran}

\usepackage{cite}
\usepackage[pagebackref=true,breaklinks=true,colorlinks,bookmarks=false]{hyperref}       % hyperlinks
\usepackage{url}            % simple URL typesetting
\usepackage{float}
\usepackage{graphicx}
\usepackage{subfig}
\usepackage{amsmath,mathrsfs}
\usepackage{amsfonts}
\usepackage{array}
\usepackage{multirow}
\begin{document}
\title{End-to-end Neural Video Coding Using a Compound Spatiotemporal Representation}
%\author{Haojie Liu\textsuperscript{\rm 1}, Ming Lu\textsuperscript{\rm 1}, Zhan Ma\textsuperscript{\rm 1}, Fan Wang\textsuperscript{\rm 2}, Zhihuang Xie\textsuperscript{\rm 2}, Xun Cao\textsuperscript{\rm 1}, and Yao Wang\textsuperscript{\rm 3} \\ \textsuperscript{\rm 1} Nanjing University, \textsuperscript{\rm 2} OPPO.Inc, \textsuperscript{\rm 3} New York University}

\author{Haojie Liu, 
        Ming Lu, Zhiqi Chen, Xun Cao,
        Zhan Ma,~\IEEEmembership{Senior Member,~IEEE}, 
        and Yao Wang,~\IEEEmembership{Fellow,~IEEE}

        }

%\author{Haojie Liu, Ming Lu, Zhan Ma, Fan Wang, Zhihuang Xie, Xun Cao, and Yao Wang}
%\textsuperscript{\rm 1} Nanjing University\\\textsuperscript{\rm 2} Horizon Robotics\\
%\author{\Large \textbf{Haojie Liu\textsuperscript{\rm 1}, Han Shen\textsuperscript{\rm 2}, Lichao Huang\textsuperscript{\rm 2}, Ming Lu\textsuperscript{\rm 1}, Tong Chen\textsuperscript{\rm 1},  Zhan Ma\textsuperscript{\rm 1}}\\ % All authors must be in the same font size and format. Use \Large and \textbf to achieve this result when breaking a line
%\textsuperscript{\rm 1} Nanjing University\\\textsuperscript{\rm 2} Horizon Robotics\\ % email address must be in roman text type, not monospace or sans serif
%}



\maketitle

\begin{abstract}
Recent years have witnessed rapid advances in learnt video coding. Most algorithms have solely relied on the vector-based motion representation and resampling (e.g., optical flow based bilinear sampling) for exploiting the inter frame redundancy. In spite of the great success of adaptive kernel-based resampling (e.g., adaptive convolutions and deformable convolutions) in video prediction for uncompressed videos, integrating such approaches with rate-distortion optimization for inter frame coding has been less successful. Recognizing that each resampling solution offers unique advantages in regions with different motion and texture characteristics, we propose a hybrid motion compensation (HMC) method that adaptively combines the predictions generated by these two approaches. Specifically, we generate a compound spatiotemporal representation (CSTR) through a recurrent information aggregation (RIA) module using information from the current and multiple past frames. We further design a one-to-many decoder pipeline to generate multiple predictions from the CSTR, including vector-based resampling, adaptive kernel-based resampling, compensation mode selection maps and texture enhancements, and combines them adaptively to achieve more accurate inter prediction.
Experiments show that our proposed inter coding system can provide better motion-compensated prediction and is more robust to occlusions and complex motions. Together with jointly trained intra coder and residual coder, the overall learnt hybrid coder yields the state-of-the-art coding efficiency in low-delay scenario, compared to the traditional H.264/AVC and H.265/HEVC, as well as recently published learning-based methods, in terms of both PSNR and MS-SSIM metrics.

   %Recent advances in video prediction have shown that feature domain alignment through feature flows can offer more accurate motion compensated predictions. Recognizing that each approach offers its unique advantage, we apply a hybrid motion compensation method that adaptively combines the predictions by both approaches. Specifically, we propose to generate the optical flow, the multiple groups of feature flows, compensation mode selection maps and texture enhancements through a single compound motion representation. This representation is derived by a recurrent information aggregation module that propagates a quantized temporal feature over multiple frames.
   %Experiments show that the proposed method can provide better prediction and is more robust to occlusions and complex motions, yielding the state-of-the-art coding efficiency in low-delay scenario, compared to the recently published learning-based methods, as well as the traditional H.264/AVC and H.265/HEVC, in terms of both PSNR and MS-SSIM metrics. 

\end{abstract}


\ifCLASSOPTIONpeerreview
\begin{center} \bfseries EDICS Category: 3-BBND \end{center}
\fi
\IEEEpeerreviewmaketitle

{\section{Introduction}\label{sec:introduction}}
% === significance ===
\IEEEPARstart{B}{uilt} upon the significant advancements of artificial neural networks (ANN), end-to-end learnt image coding algorithms~\cite{balle2018variational,minnen2018joint,chen2019neural,mattavelli2019mpeg} have made a great leap with superior coding efficiency both objectively and subjectively, over the traditional standard compliant image codecs, such as the JPEG~\cite{wallace1992jpeg}, JPEG2000~\cite{skodras2001jpeg}, BPG~\cite{bellard2015bpg} (e.g., intra profile of the High-Efficiency Video Coding [HEVC]) and even VVC (Versatile Video Coding) intra~\cite{VVC_spec}. This is mainly because spatial redundancy of a frame can be well exploited by local (e.g., vanilla convolutions) or global transforms (e.g., nonlocal operations), nonlinear activations, adaptive context modeling, and spatial-channel wise attention weighting, {\it etc.} that are integrated in entropy minimized autoencoders or variational autoencoders (VAEs). Specifically, these approaches train analysis transforms (in encoder) and synthesis transforms (in decoder) altogether with a learnt entropy  module to optimize the overall rate-distortion (R-D) efficiency.  %Recently JPEG committee has issued a Call-for-Proposal 
\begin{figure*}[t]
\centering
\includegraphics[scale=0.315]{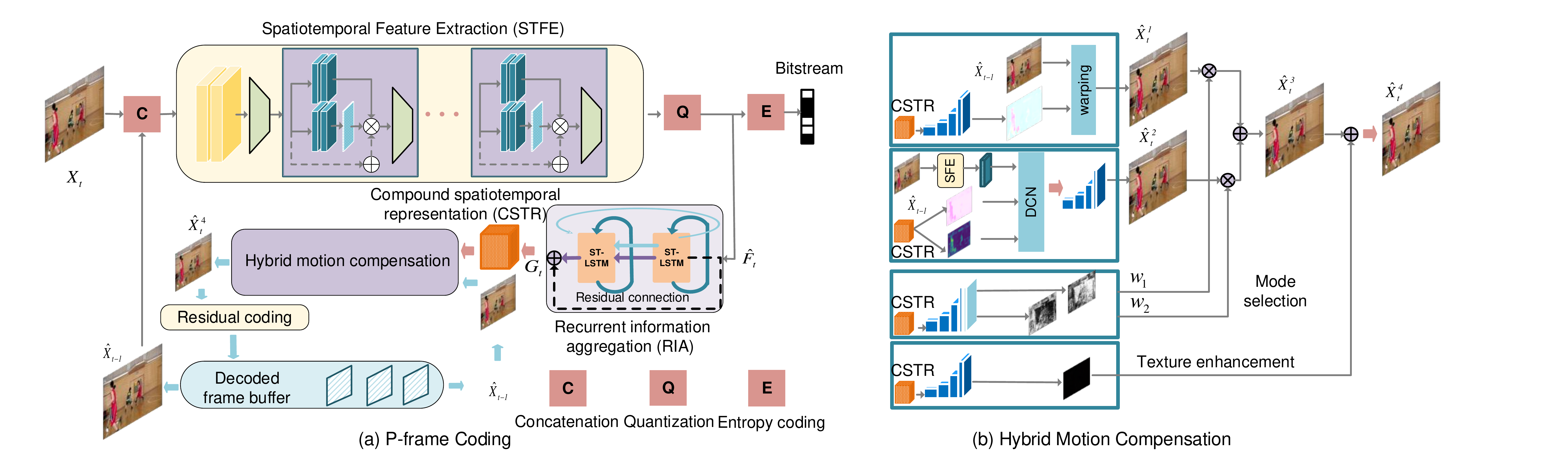}
\caption{{Hybrid Motion Compensation Using a Compound Spatiotemporal Representation.} (a) P-frame Coding: The Spatiotemporal Feature Extraction (STFE) module generates the spatiotemporal feature from the previous and the current frame, which is then quantized and entropy coded. The compressed features are recursively embedded in the ST-LSTM-based RIA to derive the CSTR for subsequent hybrid compensation; (b) Hybrid Motion Compensation: The CSTR is decoded into the vector-based motion representation, adaptive kernel-based motion representation, compensation mode weights, and texture enhancements for generating the final prediction. The SFE stands for the Spatial Feature Extraction, sharing the same network structure as the STFE by re-using the attention-based local and nonlocal modules in~\cite{chen2019neural}.}
\label{fig:inter}
\end{figure*}

Compared with aforementioned image compression methods, it is more challenging to develop an end-to-end optimized learnt video coder. Prior works can be divided into two main categories. The intuitive way is to extend the 2D autoencoder for image compression to a 3D architecture~\cite{pessoa2020end} to directly characterize spatiotemporal features in the latent space for a group of pictures (GoP). Unfortunately, such approaches so far have not provided comparable coding efficiency to the HEVC as reported in~\cite{habibian2019video}.
In practice, structural delay is also inevitable in this design which is dependent on the number of frames in a GoP.

Another promising direction is to generate explicit motion information in the form of optical flow, and then perform motion compensation to predict the frame being coded, following the classical hybrid video coding framework. In general, separate models to represent intra-pixels, inter motions and residual are jointly trained to formulate a learnt hybrid video coding system. Such approaches have yielded better coding efficiency than the HEVC compliant x265 implementation ~\cite{Lu_2019_CVPR,lin2020m,agustsson2020scale,liu2020neural}. Either pre-trained~\cite{ilg2017flownet,sun2018pwc} or unsupervised optical flow models~\cite{liu2019learned} can be applied to generate the motion information. In order to get more accurate motion compensated predictions, pre-processing network, multi-frame reference, multiscale or scale-space representation, etc, have recently been proposed~\cite{lin2020m,agustsson2020scale,liu2020neural}. Yet, only relying on the optical flow is insufficient to characterize complex temporal motion in natural scenes. Occlusions and irregular motions often lead to ghosting effects and fake textures along motion edges causing large residual signals. Furthermore, optical flow generation models are vulnerable and sensitive to quantization noise, which needs to be tuned carefully for the generalization and stability.
\begin{figure}[t]
\centering
\includegraphics[scale=0.42]{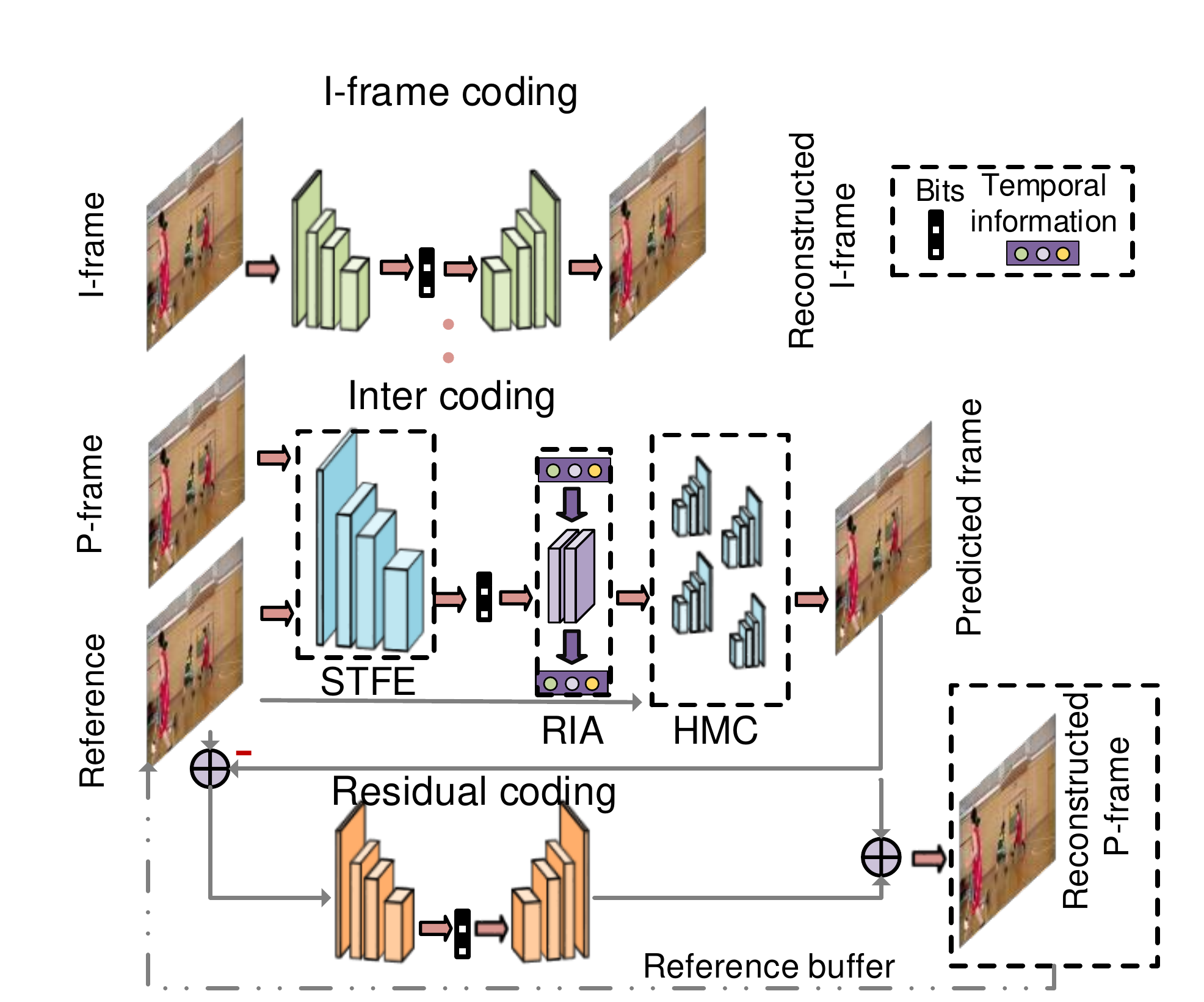}
\caption{{The Overall Compression Framework.}
The first frame is encoded using an autoencoder-based learnt image compression model. Each P-frame is described by a quantized and entropy coded spatiotemporal feature and a residual compression module. The spatiotemporal feature is used to generate the predicted frame as further detailed  in Fig.~\ref{fig:inter}.}
\label{fig:video_architecture}
\end{figure}

The proposed system uses the learnt hybrid video coding framework, but substantially modifies the motion-compensated prediction process as shown in Fig.~\ref{fig:inter}. In order to mitigate the difficulty arising from the optical flow motion representations, we  design a compound spatiotemporal representation (CSTR) from which both vector-based motion representation (i.e. optical flow), adaptive kernel-based motion representations, as well as remaining spatial information can be derived. In particular, the CSTR is decomposed by several independent sub-decoders to generate multiple predictions based on the previous decoded frame, which are then adaptively combined. This one-to-many information decomposition enables region-adaptive frame reconstruction, and will be called hybrid motion compensation (HMC).
\begin{figure*}[t]
\centering
\includegraphics[scale=0.40]{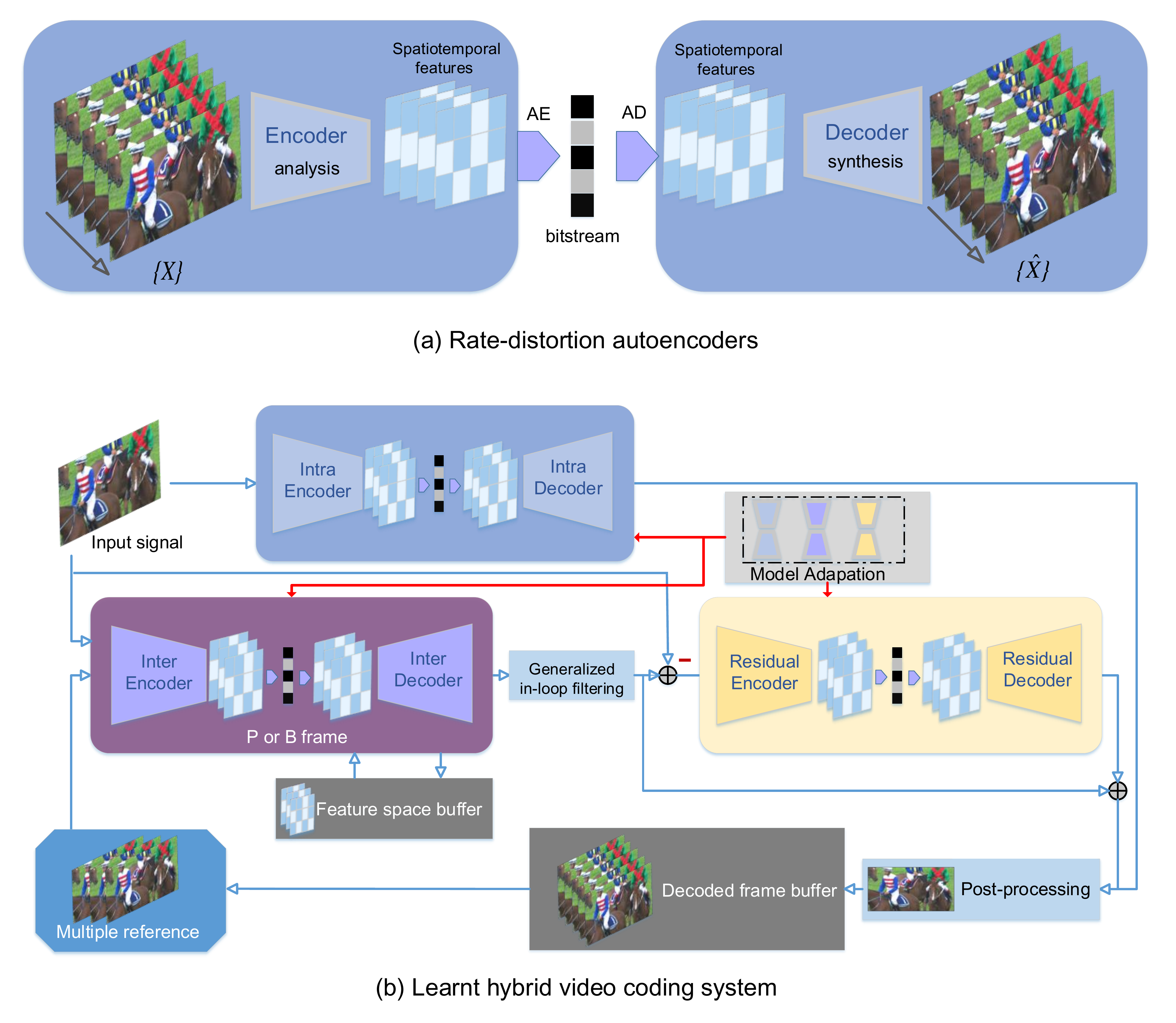}
\caption{{Illustration of two learnt video coding frameworks.} (a) Rate-distortion autoencoders: The 3D architecture is a natural extension from a learnt image coder to encode the quantized spatiotemporal features with a embedded temporal conditional entropy model. (b) Hybrid video coding: Following the traditional video coding framework, another attempt is to treat the intra-pixel, inter-motion and residual signals separately, and integrate these three coding models for joint rate-distortion optimization. }
\label{fig:main_video_coder}
\end{figure*}
Specifically, the vector-based motion representation decoder generates the compressed optical flow for bilinear warping-based compensation in the pixel domain. It works well in regions with simple motion and no occlusions. The adaptive kernel-based representation decoder offers learnable offsets and attention maps through deformable convolutions (DCN) for feature alignment. Furthermore, as feature extraction from the referenced frame can transform different motions and objects into different feature channels~\cite{Gui_2020_CVPR}, we introduce group convolutions (GC) to generate multiple groups of offsets and attention maps for different channels to effectively alleviate the artifacts introduced by occlusions and irregular motions. The compensation mode decoder generates the weights for combining the two predictions. Finally, the texture enhancement decoder generates texture information which is not captured by the motion decoders to correct the motion-compensated prediction. 

To derive the CSTR, the encoder extracts the spatiotemporal features based on the current and the previous decoded frame, which are then quantized and entropy coded. The decoded quantized features are further combined with the recursively aggregated information in a residual spatiotemporal LSTM (ST-LSTM) module to generate the CSTR. %Moreover, we also discussed the performance with different embedding ways of ST-LSTM for CSTR aggregation.

The above approach for generating the compound spatiotemporal representations and performing HMC (Fig.~\ref{fig:inter}) for inter prediction has been integrated into an end-to-end low delay neural video coding framework as shown in Fig.~\ref{fig:video_architecture}. The various encoder and decoder modules in this framework follow the network architecture utilizing attention modules proposed in~\cite{chen2019neural}.

{\bf The main contributions of this work include:}

{1)} We present the concept of hybrid motion compensation (HMC), which adaptively combines predictions obtained by different motion descriptions that are derived from the compound spatiotemporal representation (CSTR). This CSTR is further decoded by a vector-based motion decoder, an adaptive kernel-based motion decoder, a motion compensation mode decoder and a texture enhancement decoder, respectively, to enable HMC. Note that all the decoded components are generated from the compressed CSTR, and hence are rate-dependent.

{2)} We introduce a novel way of generating the CSTR, which recursively aggregates the encoder-generated quantized spatiotemporal features through a stacked residual ST-LSTM. %Three different embedding ways of ST-LSTM are discussed for the CSTR acquisition to show that we can keep the similar coding efficiency with different complexity integration as we need.

{3)} Our end-to-end optimized video coding framework does not rely on any pre-trained optical flow networks and achieves the state-of-the-art performance against the published learning-based solutions and the traditional video codecs H.264/AVC and H.265/HEVC in terms of both PSNR (Peak Sigal-to-Noise Ratio) and MS-SSIM~\cite{wang2003multiscale}.

\section{Related Work} 
\label{sec:related_work}

This section briefly reviews relevant techniques in learnt image and video compression.
\subsection{Learnt Image Compression}
 Most learnt image compression methods~\cite{balle2018variational,minnen2018joint,Cheng_2020_CVPR,chen2019neural,lee2019end} are built on variational autoencoder (VAE) structure for feature analysis and image synthesis. Nonlinear transforms, such as the generalized divisive normalization (GDN)~\cite{balle2015density}, non-local attention modules (NLAMs)~\cite{chen2019neural} and multi-frequency decomposition~\cite{akbari2020generalized}, etc, are mainly aimed at extracting more compact latent representations to reduce the entropy, while maintaining the reconstruction quality. To better model the probability distribution for entropy coding, autoregressive spatial contexts and hyperpriors are used to estimate the conditional probability of the quantized latent symbols. Furthermore, hierarchical priors~\cite{hu2020coarse} generate multi-level compressed feature representation and model the probability step-by-step. Global context~\cite{lee2019end} selects a wider range of encoded pixel regions to model the probability and improve the  entropy coding module for bit rate saving.

\subsection{Learnt Video Compression} 
%Traditional video compression technologies have been developed nearly two decades and keep following a hybrid coding system with well-designed hand-crafted engineering using variable block size. The performance is improved progressively with the development of new generation of video coding standard such as H.264, HEVC and VVC which has brought great progress in video service and applications.
%Currently learnt video coding is extended from the learnt image compression by further exploiting the inter frame redundancy. Accurate motion representation (e.g., estimation and compensation) plays a crucial role for overall coding efficiency.
Currently learnt video coding methods can be divided into two categories as show in Fig.~\ref{fig:main_video_coder}. The 3D autoencoder is a natural extension of learnt image compression to extract joint spatiotemporal features for analysis and synthesis. By explicitly exploiting the inter frame redundancy, a hybrid video coding system leverages accurate motion representation (e.g., optical flow estimation and compensation), which plays a crucial role for the overall coding efficiency.

 {\bf Non-Flow Based Motion Representation.} Chen {\it et al.}~\cite{chen2017deepcoder} made the very first attempt to establish a learnt video coding framework named DeepCoder but it still relied on the traditional block-based motion estimation and compensation for temporal motion representation. Wu {\it et al.}~\cite{wu2018vcii} later applied the recurrent neural network based video interpolation  which was then integrated with the residual coding for inter frame coding. On the other hand, Habibian {\it et al.}~\cite{habibian2019video}  utilized the rate-distortion autoencoder to directly exploit spatiotemporal redundancy in a group of pictures (GoP) with a temporal conditional entropy model, making it difficult for ultra-low-latency applications due to GoP-induced structural delay.

 {\bf Flow-based Motion Representation.} Recent studies have demonstrated that utilizing compressed optical flows to directly specify motion is an attractive solution for inter frame prediction~\cite{Lu_2019_CVPR,liu2019learned,agustsson2020scale}. For example, Lu {\it et al.}~\cite{Lu_2019_CVPR} proposed a low-latency DVC, in which a pre-trained flow estimation network FlowNet2~\cite{ilg2017flownet} was applied and the generated optical flow is subsequently compressed using a cascaded autoencoder. 
%Temporal error control and adaptation~\cite{lu2020content} is studied to minimize the error propagation and improve the performance of DVC. More extensions such as model complexity analysis and variable quantization can be found in~\cite{lu2020end}. 
Though encouraging coding efficiency was reported, solely relying on (the nearest) previous decoded frame for flow derivation was insufficient to capture the complex and irregular motion in nature scene and resist the quantization error. Thus, multiple reference frame-based long-term prediction was suggested in MLVC~\cite{lin2020m} by caching more previously decoded frames and motion fields for rate-distortion optimization. Besides, bi-directional motion was studied in~\cite{djelouah2019neural, Yang_2020_CVPR} by additionally exploring the ``future frames". Both long-term and bi-directional predictions had attempted to better characterize complex motion to improve the coding efficiency, but at the expense of more memory consumption and larger coding latency. 

In general, all of the previous methods still focused on the single optical flow for motion compensation which can be defined as a vector-based resampling and applied in the pixel domain. Agustsson {\it et al.}~\cite{agustsson2020scale} further presented a scale-space flow generation and trilinear warping method for motion compensation, in which an additional scale field is estimated, which enabled the frame prediction from a blurred video frame when the actual motion is complex and the estimated optical flow is inaccurate, leading to a reduced prediction residual. Liu {\it et al.}~\cite{liu2020neural} used pyramid optical flow decoder for multi-scale compressed optical flow estimation and applied a progressive refinement strategy with joint feature and pixel domain motion compensation. Both methods in~\cite{agustsson2020scale,liu2020neural} had shown noticeable coding gains of inter frames. It is worth to point out that these methods are not mutually exclusive and may be combined to achieve additional gains.%additive gains may be captured by combining the long-term/bi-directional prediction, scale-space flow and multi-scale flow representation.  

\begin{figure*}[t]
\centering
\includegraphics[scale=0.237]{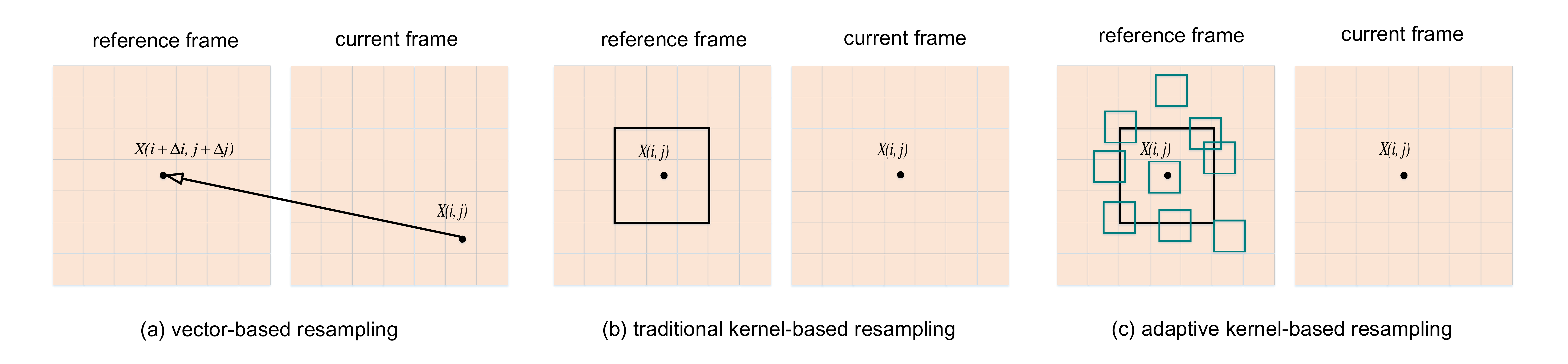}
\caption{{Different Resampling Solutions for Motion Compensated Prediction.}}
\label{fig:resampling}
\end{figure*}

{\bf Vector-based Resampling \& Kernel-based Resampling.}
Although adaptive-kernel based resampling is widely used and discussed in video interpolation, extrapolation and processing, it has not been explored so far for learnt video coding under the entropy minimization constraints. Given two consecutive frames, vector-based resampling uses a explicit transformation with bilinear interpolation to take the frame $t$ to frame $t+1$ as show in Fig.~\ref{fig:resampling}. This transformation can also be defined as a kernel-based sampling when changing the bilinear sampling into vanilla convolutions as shown in Fig~\ref{fig:resampling} (b). However, these traditional convolution operations are limited to its spatial invariance to model the complex and long-term temporal transformations. Adaptive kernel-based resampling (e.g., deformable convolutions) further address the issues based on the spatiotemporal variable receptive fields.

{\textbf {Temporal Correlation Exploration in both Pixel and Feature Domain.}} Conventional video coding mostly utilized the previous decoded frames (in pixel domain) for exploiting the temporal correlation. Leveraging the recurrent network models, learnt video coding could exploit correlations in both pixel and feature domains over multiple frames. For instance, Rippel {\it et al.}~\cite{rippel2019learned} maintained a recurrent state for multiple frame information fusion and jointly decoded the optical flow and residual information for end-to-end learning.  Another example was to aggregate features of previous reconstructed motions and residuals as temporal priors for conditional probability estimation for the current motion and residual in inter coding~\cite{liu2019learned,yang2020learning}.  

\begin{figure}[t]
\centering
\includegraphics[scale=0.35]{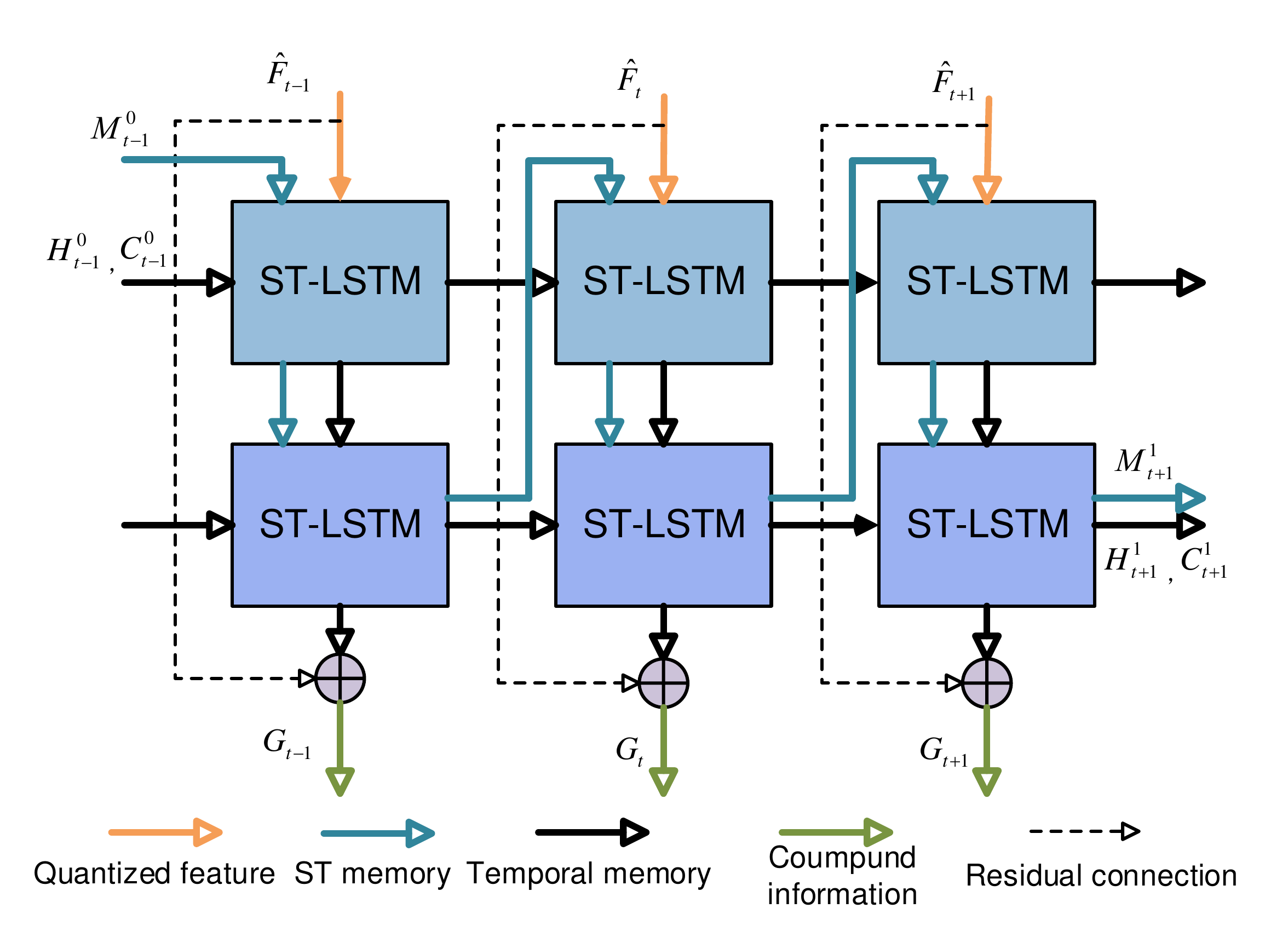}
\caption{{Information Flow in the residual ST-LSTM.}}
\label{fig:st_lstm}
\end{figure}

\section{Proposed Video Coding System} \label{sec:nvc}
%\subsection{An Overview of Neural Video Coding}

Figure~\ref{fig:inter} provides an overview of the proposed hybrid motion compensation framework using the compound spatiotemporal representation (CSTR) for P-frame compression. This CSTR is generated by aggregating compressed latent spatiotemporal features recurrently using the  recurrent information aggregation module (see in Sec.~\ref{sec:ria}) for efficient information fusion and embedding. The hybrid motion compensation (HMC) module implements multiple independent decoders to parse the CSTR for different motion descriptions that will be used to generate different motion-compensated predictions, which are then adaptively combined to generate the final prediction frame (see Sec.~\ref{sec:hmc}).

\subsection{Recurrent Information Aggregation} \label{sec:ria}
\begin{figure}[t]
\centering
\includegraphics[scale=0.38]{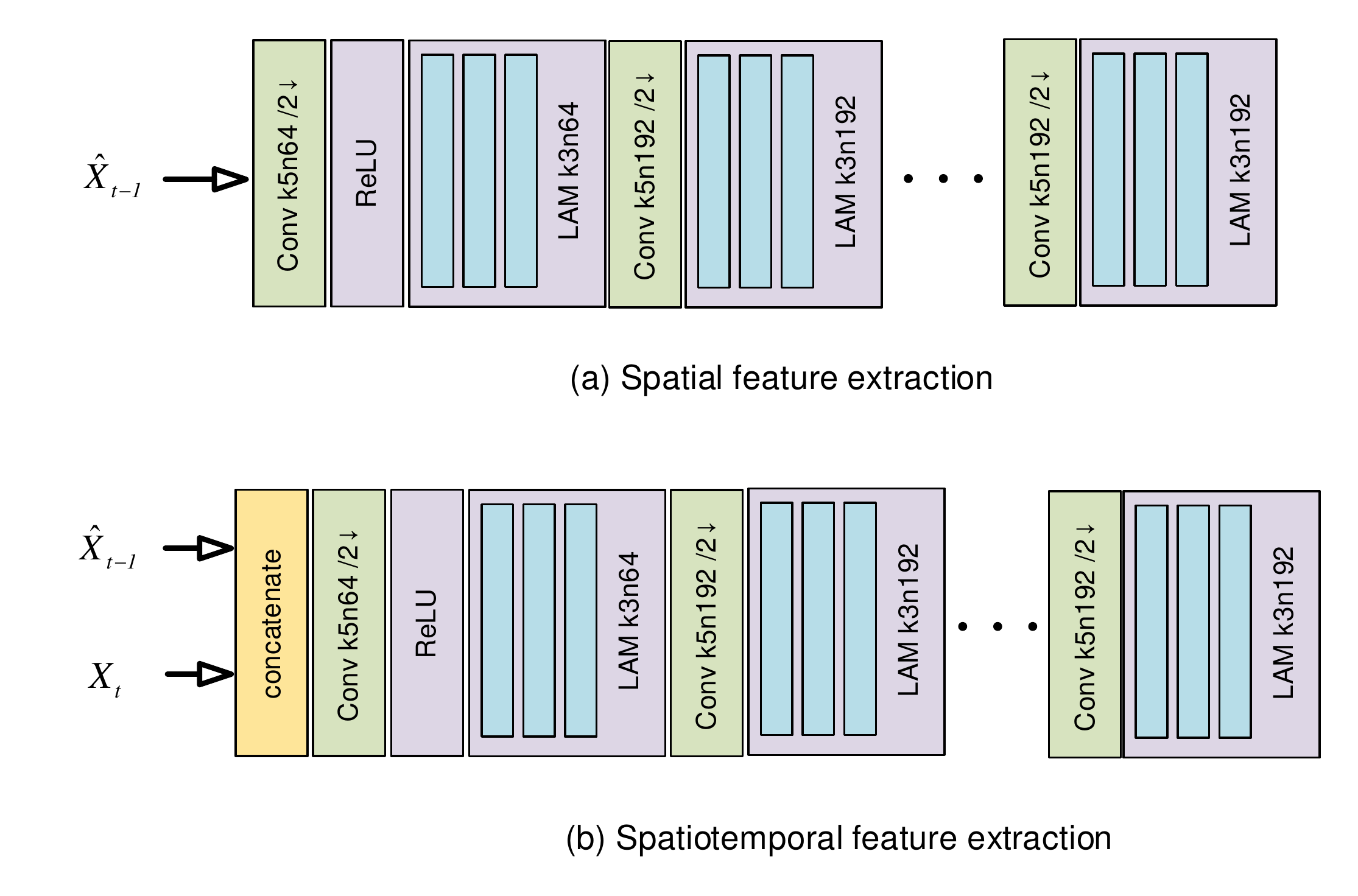}
\caption{{Feature Extraction.} (a) Spatial feature extraction (SFE) is used to generate the referenced features from the former decoded frame for adaptive kernel-based motion compensation in the feature domain. (b) Spatiotemporal feature extraction (STFE) takes the past decoded frame and the current frame as inputs to obtain the fused spatiotemporal features $\mathcal{ \hat{F}}_t$. These features are quantized and entropy coded for the bit streams. They are further decoded and fused in the recurrent information aggregation to obtain the CSTR. Here, "k3n64 2↓" means 3$\times$3 convolutions with 64 channels and a downsampling factor of 2. }

\label{fig:sfe}
\end{figure}

We employ stacked local and nonlocal attention modules with 4 2$\times$2 downsampling layers as a spatiotemporal feature extraction (STFE) module detailed in Fig.~\ref{fig:sfe}(b). 
Then quantization is applied to obtain the latent spatiotemporal features.
Our proposed recurrent information aggregation (RIA) module utilizes stacked ST-LSTM~\cite{wang2017predrnn} to combine  currently decoded  latent spatiotemporal features with the past information accumulated up to the last decoded frame for the derivation of a CSTR as shown in Fig.~\ref{fig:st_lstm}. Compared to the traditional ConvLSTM, ST-LSTM adds a spatiotemporal memory state and an additional information flow. The spatiotemporal state is propagated not only horizontally (i.e. temporally), as with the ConvLSTM, but also vertically (i.e. spatially) across multiple representation levels.

 With the two-layer recurrent model exemplified in Fig.~\ref{fig:st_lstm}, given the decoded quantized features $\mathcal{\hat {F}}_t$, the CSTR ${G}_t$ is obtained as follows: 
\begin{align}
  \mathcal{H}_{t}^0,\mathcal{C}_{t}^0, \mathcal{M}_{t}^0 &= \textrm { ST-LSTM}_t^0(\mathcal{\hat {F}}_t,\mathcal{H}_{t-1}^0,\mathcal{C}_{t-1}^0,\mathcal{M}_{t-1}^1), \label{eq:AIA1}\\
  \mathcal{H}_{t}^1, \mathcal{C}_{t}^1,\mathcal{M}_{t}^1 & = \textrm{ST-LSTM}_t^1(\mathcal{H}_t^0,\mathcal{H}_{t-1}^1,\mathcal{C}_{t-1}^1,\mathcal{M}_{t}^0), \label{eq:AIA2}\\
   {G}_t &= \mathcal{\hat {F}}_t+\mathcal{H}_{t}^1. \label{eq:AIA3}
\end{align}
$\mathcal{M}_t^l$ denotes the spatiotemporal memory state at frame $t$  and layer $l$ in the recurrent network. Note that $\mathcal M_t^0 $ depends on the highest level memory state of the previous frame $\mathcal M_{t-1}^1$; while $\mathcal M_t^1$ depends on the lower level state at the current frame $\mathcal M_t^0$. $\mathcal{C}_{t}^l$ is a temporal memory state and controls the temporal information flow. $\mathcal{H}_{t}^l$ maintains the previously-accumulated decoded information both spatially and temporally. Different from the ST-LSTM introduced in~\cite{wang2017predrnn}, we introduce residual connections across the recurrent layers for faster convergence in generating the final CSTR ${G}_t$. 

%-------------------------------------------------------------------------
\subsection{ Hybrid Motion Compensation} \label{sec:hmc}
Prior methods usually rely on the vector-based resampling to directly remove the temporal redundancy between two consecutive frames. The past decoded frame and the current frame are fed into the pre-trained flow networks for further compression or end-to-end optimized with bottleneck layer rate constraints without supervision. The quantized latent features from the bottleneck layer can be taken as simple motion representations and then be decoded into the vector-based motions for motion compensation. Such methods often fail in regions with occlusions and fast motions because of the limitation of the bilinear warping in the pixel domain. The scale-space flow introduced in~\cite{agustsson2020scale} allows the encoder to switch to a blurred prediction in regions where the estimated optical flow may be wrong, to minimize the rate needed to code the prediction residual. We take a significant step further, by introducing hybrid motion compensation (HMC), which adaptively combines multiple predictions. 
 Two independent sub-networks produce two predictions from the compound spatiotemporal representation ${G}_t$ generated by the RIA module. Furthermore, a compensation mode decoder generates the weight map from ${G}_t$, which indicates the spatial varying weights for combining the two predictions. A fourth component is also decoded from ${G}_t$, which is the texture enhancement to be added to the weighted prediction for spatial information compensation.

{\textbf{Vector-based motion decoder $\mathcal{D}_f$}.} 
$\mathcal{D}_f$ uses an attention-based network architecture to generate the backward optical flow field ${f} = ({f}_x,{f}_y)$ from the compound representation ${G}_t$ for direct vector-based resampling in pixel domain. The backward bilinear warping is used to warp the previous decoded frame $\hat{X}_{t-1}$ by:

\begin{align}
f &= \mathcal{D}_f({G}_t),\\
  {\hat{X}_{t}^1} &= \mathbf{Warping}({ \hat{X}_{t-1}}, f), \label{eq:warping}
\end{align}

{\textbf{Adaptive kernel-based motion decoder with occlusion aware attention $\mathcal{D}_m$}.} Adaptive kernels (e.g., dynamic filters and deformable convolutions) have been used successfully recently in video prediction and processing for effective feature alignment~\cite{dai2017deformable,zhu2019deformable,Gui_2020_CVPR}. Here, we incorporate the attention mechanism proposed in~\cite{Gui_2020_CVPR} to enable the spatially varying activation to handle occluded/disoccluded regions. We firstly extract the reference features $\hat{X}_F$ as the feature domain reference from the decoded reference frame $ \hat{X}_{t-1}$ using a spatial feature extraction module as shown in Fig.~\ref{fig:sfe}(a). Assuming that different channels of $\hat{X}_F$ in the latent space represent different objects and background after the complicated transformation and mapping in~\cite{Gui_2020_CVPR}, we also introduce group convolutions to generate multiple learnable offsets and attention maps for different feature groups ${G}_t$. Empirically, we  divide all $N=192$ feature maps into 4 groups to provide a good tradeoff between the prediction accuracy and the complexity. The detailed pipeline can be denoted using:
\begin{align}
  {\hat{X}_t^2} = \mathcal{D}_m({\hat{X}_F},{G}_t), \label{eq:deformable}
\end{align}
We have found that the prediction generated using adaptive kernel-based motion decoder is more stable and robust to fast or irregular motions.
 Occlusion-aware attention maps in deformable convolutions can also handle the prediction of occluded areas effectively compared with the vector-based motion compensation in pixel level.

\begin{figure*}[t]
\centering
\includegraphics[scale=0.180]{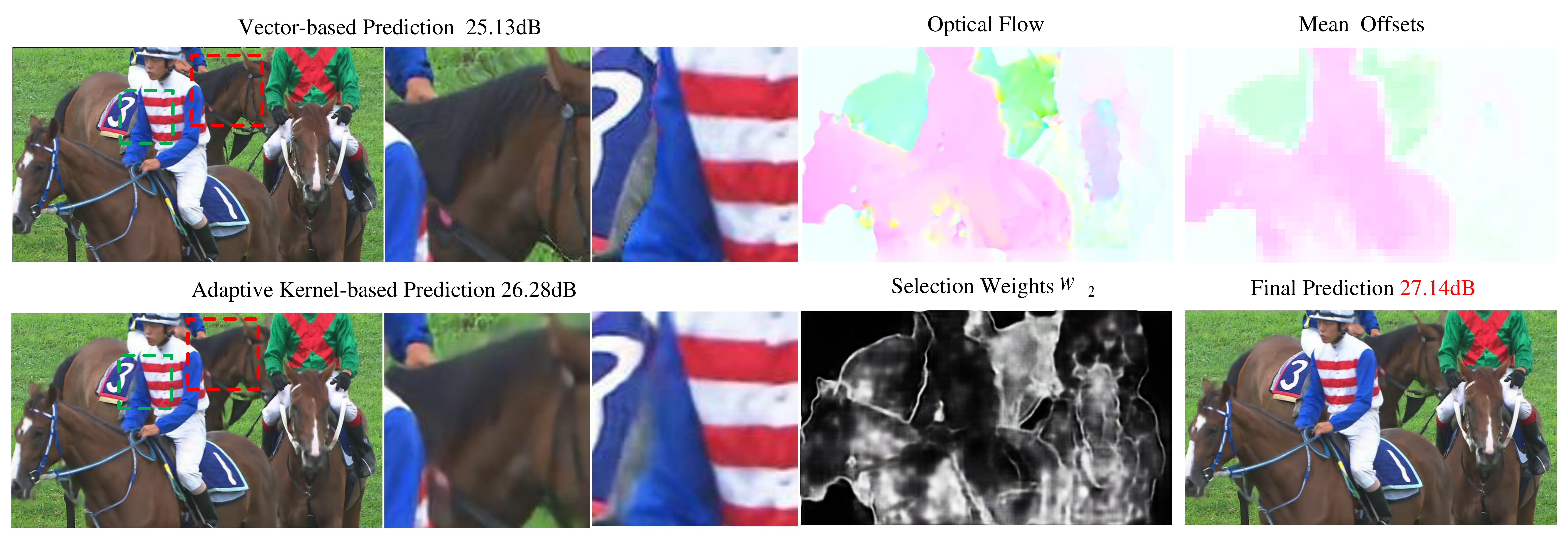}
\caption{{Visualization of the Internals of Our Inter Coding Framework.} Under the rate constraint of the compound representation, the independent decoders can generate multiple inter predictions. Vector-based prediction works well on stationary areas and slow motions as shown in the background. On the other hand, adaptive kernel-based prediction generates less artifacts along moving edges and regions undergoing fast motion and experiencing motion blur.}

\label{fig:different_decoders}
\end{figure*}

{\textbf{Compensation mode decoder $\mathcal{D}_w$}.} To benefit from the advantages of both the vector-based prediction $\hat{X}_t^1$ and adaptive kernel-based prediction $\hat{X}_t^2$, a compensation mode decoder $\mathcal{D}_w$ is developed to generate soft weights $w_1$ and $w_2$ for the linear combination from $ \hat{X}_t^1$ and $\hat{X}_t^2$, leading to the combined prediction:
\begin{align}
  {\hat{X}_t^3} = {w_1}\cdot{ \hat{X}_t^1}+{w_2}\cdot{\hat{X}_t^2}, \label{eq:weighted prediction}
\end{align}
Note that the weights are the same size as the reference frame $\hat{X}_{t-1}$. And the $\mathrm{sigmoid}$ function is used to keep the values in both $w_1$ and $ w_2$  between 0 to 1. As shown in Fig.~\ref{fig:different_decoders}, adaptive kernel-based prediction is allocated more attention in noisy areas and areas with occlusion due to motion (e.g. along the moving edges). We note that this weighted prediction is more general than the single vector-based prediction or adaptive kernel-based prediction. Specifically, when $ w_2$ = 0, the prediction only uses $\hat{X}_t^1$ and is similar to the bilinear warping based approaches. When $w_1$ = 0, the prediction only relies on $\hat{X}_t^2$ with rate constrained offsets and attention maps for adaptive kernel-based prediction.

\begin{figure*}[t]
\centering
\includegraphics[scale=0.365]{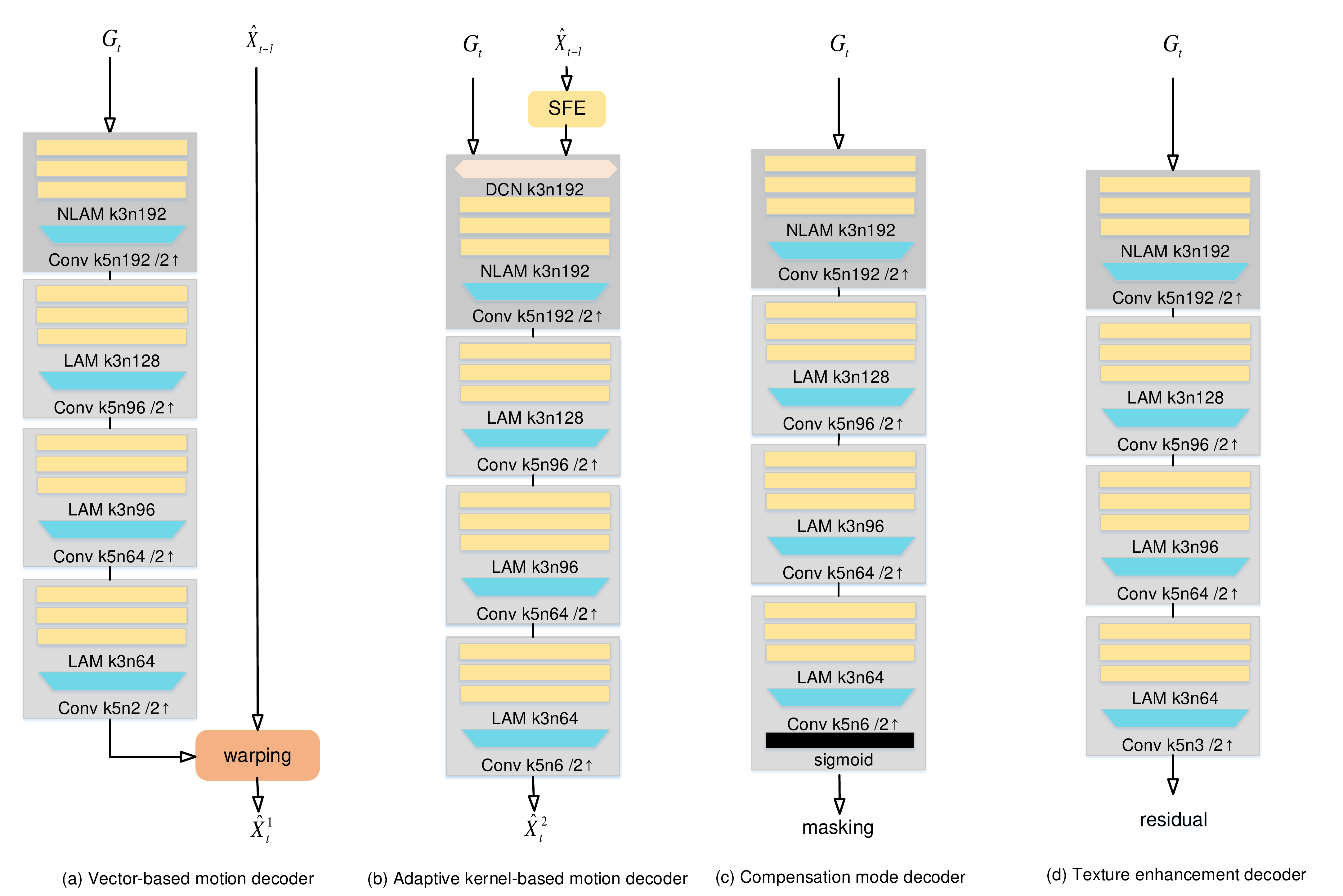}
\caption{{Network Architecture for
the Independent Decoders.} These sub-networks are used for decoding different motion descriptions and mode selection masks and the residual texture information to enable the hybrid motion compensation. Here, LAM and NLAM represents the attention based local and nonlocal modules in~\cite{chen2019neural}. "k3n64" means that convolutions in LAM and NLAM use 3$\times$3 strides with 64 channels.}
\label{fig:decoders}
\end{figure*}

{\textbf{Texture enhancement decoder $\mathcal{D}_t$}.} We want to further extract the remaining texture information contained in ${G}_t$ but are not captured by the predicted frame $\hat{X}_t^3$ using the decoded motion information and mode selection maps. Because we want the CSTR to be mainly driven by the vector-based motion decoder and the adaptive kernel-based decoder with their linear combination using spatial-varying weights, we apply a gradient stop setting when training this sub-network, so that the gradient associated with the loss due to this component does not backpropagate beyond the CSTR. The final prediction $\hat{X}_t^4$ is acquired by:

\begin{align}
  {\hat{X}_t^4} = { \hat{X}_t^3}+\mathcal{D}_t(\mathrm{stop\_gradient}({G}_t)), \label{eq:texture}
\end{align}

{\textbf{Network architectures for these independent decoders}.} The details of the independent decoders with specific output channels $N$ are shown in Fig.~\ref{fig:decoders}. For the vector-based motion decoder $\mathcal{D}_f$, $N$ is set to 2 to generate the optical flow for pixel domain motion compensation. $\mathcal{D}_m$ firstly utilized the ${G}_t$ to obtain the groups of offsets and attention maps that are then applied on the referenced features from $\hat{X}_{t-1}$. The aligned features are then synthesized through the following decoding layers with multiple stage upsampling to generate the adaptive kernel-based prediction. $\mathcal{D}_w$ output totally 6 channels to provide 3 channel weighted maps for $\hat{X}_t^1$ and $ \hat{X}_t^2$, respectively. A sigmoid layer is applied to control the range of pixel activation. $\mathcal{D}_t$ output 3 channels to estimate the remaining texture for enhancement. Here, the independent decoders share the similar core network architecture as shown in Fig.~\ref{fig:decoders} with some additional layers to generate the specific predictions.

% 2021-6-26

{\textbf{Joint rate-distortion optimization with multiple sub-networks}.} We optimize the whole inter prediction system using a joint rate-distortion loss function:
\begin{align}
  \mathcal{L}_t^p = \lambda{\sum_{i=1}^{4}}d_1( \hat{X}_t^i,{X}_t)+H(\mathcal{\hat F}_t) , \label{eq:joint inter}
\end{align}
where $d_1(.)$ is the $l_1$ loss and ${H(.)}$ is an entropy estimator. Note that each prediction ${ \hat{X}_t^i}$ depends on its own decoder and the shared CSTR ${G}_t$, which in turn depends on the RIA module and $\mathcal{\hat F}_t$. We use a relatively large $\lambda$ during this pre-training for fast convergence. Once the four decoders converge using this loss, the entire inter prediction network is integrated into the whole video compression system for an end-to-end optimization.

%$\lambda$ starts out relatively large for better convergence and then the inter prediction network is integrated into the whole video compression system for an end-to-end optimization.

%\subsection{Random Noise Regularization}

%In the end-to-end video coding framework, the spatiotemporal features for inter prediction and the latent feature for describing the prediction residual both have to be quantized. During network training, such quantization is emulated by adding a uniform noise~\cite{balle2016end} to enable gradient back propagation. The mismatch between the uniform noise added during the  training stage and the actual quantization error distribution during the inference stage tends to lead to significant performance drop at the inference stage, because the quantized features in one frame can affect many subsequent frames. 

%To overcome the sensitivity of the trained network to such distribution mismatch, we introduce random noise regularization (RNR) during the training stage to perturb the decoded features. We add a random noise with a mixed uniform distribution in the range of  $-{2^{n}}$ to ${2^{-n}}$ with $n$ sampled randomly between -1 and 1 during training ($n$ is fixed within each training batch). We have found that this regularization strategy leads to more stable P-frame prediction and residual coding.

\subsection{The Overall Video Compression Model}

The overall framework of video compression follows the low-latency coding pipeline without bi-directional prediction, as shown in Fig.~\ref{fig:video_architecture}. Given a group of frames $ X_0, X_1,..., X_t$, the first frame (I-frame) $X_0$ is encoded using a learnt image compression network in~\cite{chen2019neural}, which generates the encoded quantized latent features $ \hat {Y}_0$ , from which the decoder generate the reconstructed frame $\hat{X}_0$. Then the following frame (P-frame) $X_t$ and the previous decoded frame $\hat{X}_{t-1}$ are fed into the STFE to generate the feature $\mathcal{F}_t$, which is then quantized and entropy coded. The (RIA) unit  combines the quantized representation $\mathcal{\hat F}_t$ with the accumulated temporal information stored in  the ST-LSTM network to generate a compound spatiotemporal representation ${G}_t$. Subsequent independent modules $\mathcal{D}_f,\mathcal{D}_m,\mathcal{D}_w$ and $\mathcal{D}_t$ use ${G}_t$ and the previous decoded frame to generate a final prediction $\hat{X}_t^4$. A learnt residual image compression network (following the same structure as~\cite{chen2019neural}) is then applied to encode the prediction residual $ R_t =  X_t -  \hat{ X}_t^4$ into a representation $ \hat{Y}_t^r$ and obtain the final residual compensated results by $ \hat{X}_t =  \hat{X}_t^4 +  \hat{R}_t$. Here, all the quantized latent representations $ \hat {Y}_0$, $\mathcal{\hat F}_t$ and $ \hat {Y}_t^r$ are entropy-coded using the joint priors estimation method~\cite{minnen2018joint} to model the probability distributions of the latent variables. 

We first pre-train the various sub-networks in the inter-coding framework using the loss function in Eq.~(\ref{eq:joint inter}). Here, we use a relatively large $\lambda$ to achieve fast network convergence. The overall system is then optimized using a total loss for $N$ training frames:
\begin{align}
  \mathcal{L}_{total} &= {{\lambda_1}d_2( \hat{X}_0, X_0)}+{H}({ \hat {Y}_0})+{\lambda_2}{{\sum_{t=1}^{N-1}}{\sum_{i=1}^{4}}d_1( \hat{X}_t^i, X_t)}\nonumber\\ 
  &+{\lambda_3}{{\sum_{t=1}^{N-1}}{d_2}( \hat{X}_t, X_t)}+{\sum_{t=1}^{N-1}}H(\mathcal{\hat F}_t)+H( \hat{Y}_t^r) , \label{eq:joint prediction}
\end{align}
where the reconstruction loss $d_2(.)$ is either the mean square (MSE) or the negative multi-scale structure similarity (MS-SSIM). The hyper parameters $\lambda_1$, $\lambda_2$ and $\lambda_3$ control the final bit allocation among the intra, inter-motion, and inter-residual information. For the PSNR optimized models, to obtain different target rates, we set ($\lambda_1$, $\lambda_2$, $\lambda_3$) to (12800, 32, 3200), (6400, 16, 1600), (3200, 8, 800) and (1600, 4, 400). For MS-SSIM optimized models, they are (128, 32, 48), (64, 16, 24), (32, 8, 12) and (16, 4, 6). Note that even if we use a small $N$ during the training, the resulting models for inter coding and residual coding can be used for all P-frames within a group of pictures with more than $N$ frames.
 
\section{Experimental Studies} \label{sec:exp}
\subsection{Dataset and Experiment Setup}
\textbf{Training Datasets.} The intra frame compression models are trained on CLIC and DIV2K with cropped size of 256$\times$256$\times$3. Then the overall video compression models are end-to-end optimized using the Vimeo 90k~\cite{xue2019video} with cropped sample size of 192$\times$192$\times$3 pixels. We first enroll $N$ = 2 frames to pre-train the modules in P-frame coding and finally enroll $N$ = 5 frames to optimize the whole system. Optimizing the performance over multiple P frames  can make the trained network to learn to be robust to reconstruction error propagation. Our experiments have shown that training using $N=5$ frames (1 I frame followed by 4 P-frames) provides a good tradeoff between the rate-distortion performance and training complexity. 

%6-27 2:28

\textbf{Evaluation Data and Criterion.} Our method is evaluated on ultra video group (UVG)~\cite{10.1145/3339825.3394937}, MCL-JCV~\cite{7532610} and JCT-VC dataset~\cite{jct}.  Ultra video group (UVG) contains high resolution videos which are widely used for evaluating video coding and processing performance. MCL-JCV is a jnd-based
H.264/AVC video quality assessment dataset and contains several 1080p videos collected from YouTube. JCT-VC is a video collection having multiple spatial resolutions, frame rates and motion characteristics, which is usually treated as a common test dataset for traditional video coding. We set the same low-delay coding mode and GoP length (12 for UVG and MCL-JCV, 10 for JCT-VC) as DVC~\cite{Lu_2019_CVPR} and DVC\_pro~\cite{lu2020end}. Since most published works adopt sRGB domain for evaluation, we keep the same setting for fair comparison although there are some losses due to color space change from YUV format to sRGB for the raw video data.

\begin{figure*}[t]
\centering
\includegraphics[scale=0.258]{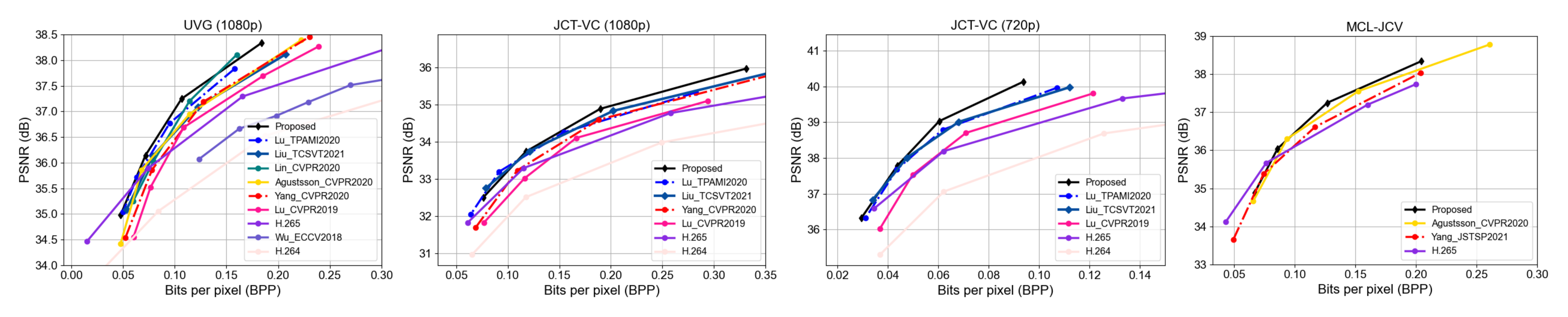}
\caption{{Rate-distortion Performance in terms of PSNR.}}
\label{fig:rd_psnr}
\end{figure*}

\begin{figure*}[t]
\centering
\includegraphics[scale=0.26]{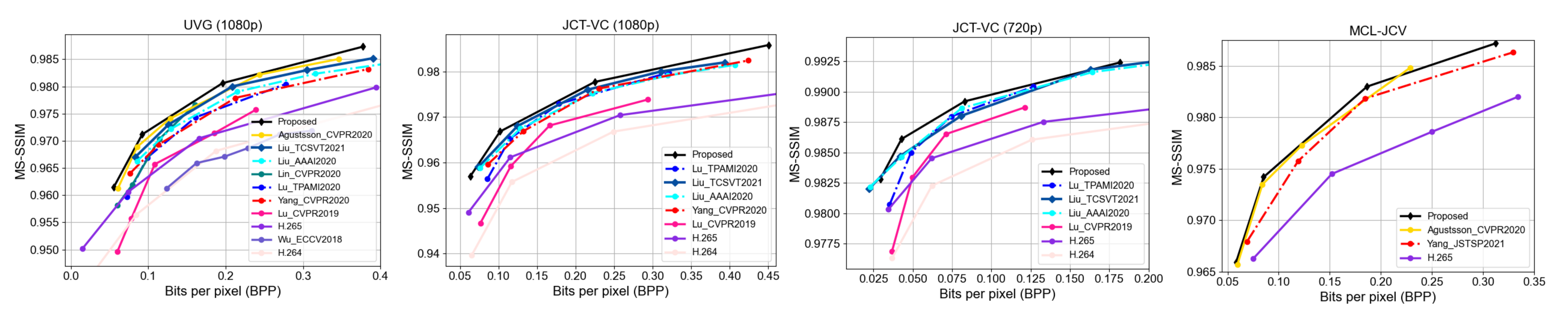}
\caption{ {Rate-distortion Performance in terms of MS-SSIM.}}
\label{fig:rd_ssim}
\end{figure*}

\begin{table*}[t]
  \centering
  \caption{BD-Rate Gains of Proposed method, NVC, HEVC, DVC and DVC\_pro against the H.264/AVC measured by PSNR on JCT-VC and UVG. }
%  \toprule[2pt]
  \begin{tabular}{c|c|c|c|c|c|c|c|c|c|c|c|c|c|c|c|c|c|c|c|c}
    %\hline
    \hline
    \multirow{3}{*}{\textbf{Sequences}}&\multicolumn{4}{c|}{\textbf{H.265/HEVC}}&\multicolumn{4}{c|}{\textbf{DVC}}&\multicolumn{4}{c|}{\textbf{NVC}}&\multicolumn{4}{c|}{\textbf{DVC\_pro}}&\multicolumn{4}{c}{\textbf{Proposed}}\\
     
    \cline{2-21}
      & \multicolumn{4}{c|}{PSNR}  &\multicolumn{4}{c|}{PSNR} & \multicolumn{4}{c|}{PSNR} & \multicolumn{4}{c|}{PSNR}& \multicolumn{4}{c}{PSNR} \\
      \cline{2-21}
    &  \multicolumn{2}{c|}{BDBR} & \multicolumn{2}{c|}{BD-(D)} & \multicolumn{2}{c|}{BDBR} & \multicolumn{2}{c|}{BD-(D)} & \multicolumn{2}{c|}{BDBR} & \multicolumn{2}{c|} {BD-(D)} & \multicolumn{2}{c|}{BDBR} & \multicolumn{2}{c|} {BD-(D)} & \multicolumn{2}{c|}{BDBR} & \multicolumn{2}{c} {BD-(D)} \\
    
     \hline
     
    {ClassB} &  \multicolumn{2}{c|}{-32.03\%} & \multicolumn{2}{c|}{0.78} & \multicolumn{2}{c|}{-27.92\%} & \multicolumn{2}{c|}{0.72} & \multicolumn{2}{c|}{-45.66\%} & \multicolumn{2}{c|} {1.21} & \multicolumn{2}{c|}{-43.66\%} & \multicolumn{2}{c|} {1.18} & \multicolumn{2}{c|}{\textbf{-47.40\%}} & \multicolumn{2}{c} {\textbf{1.32}} \\
    \hline
     
    {ClassC} &  \multicolumn{2}{c|}{-20.88\%} & \multicolumn{2}{c|}{0.91} & \multicolumn{2}{c|}{-3.53\%} & \multicolumn{2}{c|}{0.13} & \multicolumn{2}{c|}{-17.82\%} & \multicolumn{2}{c|} {0.73} & \multicolumn{2}{c|}{\textbf{-23.81\%}} & \multicolumn{2}{c|} {\textbf{1.07}} & \multicolumn{2}{c|}{-12.54\%} & \multicolumn{2}{c} {0.53} \\
    \hline
     
    {ClassD} &  \multicolumn{2}{c|}{-12.39\%} & \multicolumn{2}{c|}{0.57} & \multicolumn{2}{c|}{-6.20\%} & \multicolumn{2}{c|}{0.26} & \multicolumn{2}{c|}{-15.53\%} & \multicolumn{2}{c|} {0.70} & \multicolumn{2}{c|}{\textbf{-24.35\%}} & \multicolumn{2}{c|} {\textbf{1.21}} & \multicolumn{2}{c|}{-0.12\%} & \multicolumn{2}{c} {0.01} \\
    \hline
     
    {ClassE} &  \multicolumn{2}{c|}{-36.45\%} & \multicolumn{2}{c|}{0.99} & \multicolumn{2}{c|}{-35.94\%} & \multicolumn{2}{c|}{1.17} & \multicolumn{2}{c|}{-49.81\%} & \multicolumn{2}{c|} {1.70} & \multicolumn{2}{c|}{-48.24\%} & \multicolumn{2}{c|} {1.72} & \multicolumn{2}{c|}{\textbf{-52.56\%}} & \multicolumn{2}{c} {\textbf{2.00}} \\
    \hline
     
    {UVG} &  \multicolumn{2}{c|}{-48.53\%} & \multicolumn{2}{c|}{1.00} & \multicolumn{2}{c|}{-37.74\%} & \multicolumn{2}{c|}{1.00} & \multicolumn{2}{c|}{-48.91\%} & \multicolumn{2}{c|} {1.24} & \multicolumn{2}{c|}{-52.65\%} & \multicolumn{2}{c|} {1.37} & \multicolumn{2}{c|}{\textbf{-55.61\%}} & \multicolumn{2}{c} {\textbf{1.55}} \\

    \hline
    %\bottomrule[2pt]
  \end{tabular}
  \label{tab:BDrate}
\end{table*}
%%%%%%%%%%%%%%%%%%%%%%%%%%%%%%%%%%%%%%%%%%%%%%%%%%%%%%%%%%%%%%%%%%%%%%%%%%%%%%%%%%%%%%%%%%%%%%%%%%%%%%%%%%%%%%%%%%%%%%%%%%%%%%%%%%%%%%%%%%%%%%%%%%%%%%%%%%%%%%%%%%%%%%%%%%%%%%%
\begin{table*}[t]
  \centering
  \caption{BD-Rate Gains of Proposed method, NVC, HEVC, DVC and DVC\_pro against the H.264/AVC measured by MS-SSIM on JCT-VC and UVG. }
%  \toprule[2pt]
  \begin{tabular}{c|c|c|c|c|c|c|c|c|c|c|c|c|c|c|c|c|c|c|c|c|c|c|c|c}
    %\hline
    \hline
    \multirow{3}{*}{\textbf{Sequences}}&\multicolumn{4}{c|}{\textbf{H.265/HEVC}}&\multicolumn{4}{c|}{\textbf{DVC}}&\multicolumn{4}{c|}{\textbf{NVC}}&\multicolumn{4}{c|}{\textbf{DVC\_pro}}&\multicolumn{4}{c|}{\textbf{DVC(MS-SSIM)}}&\multicolumn{4}{c}{\textbf{Proposed}}\\
     
    \cline{2-25}
      & \multicolumn{4}{c|}{MS-SSIM}  &\multicolumn{4}{c|}{MS-SSIM} & \multicolumn{4}{c|}{MS-SSIM} & \multicolumn{4}{c|}{MS-SSIM}& \multicolumn{4}{c|}{MS-SSIM}& \multicolumn{4}{c}{MS-SSIM} \\
      \cline{2-25}
    &  \multicolumn{2}{c|}{BDBR} & \multicolumn{2}{c|}{BD-(D)} & \multicolumn{2}{c|}{BDBR} & \multicolumn{2}{c|}{BD-(D)} & \multicolumn{2}{c|}{BDBR} & \multicolumn{2}{c|} {BD-(D)} & \multicolumn{2}{c|}{BDBR} & \multicolumn{2}{c|} {BD-(D)} & \multicolumn{2}{c|}{BDBR} & \multicolumn{2}{c|} {BD-(D)} & \multicolumn{2}{c|}{BDBR} & \multicolumn{2}{c} {BD-(D)}\\
    
     \hline
     
    {ClassB} &  \multicolumn{2}{c|}{-27.67\%} & \multicolumn{2}{c|}{0.0046} & \multicolumn{2}{c|}{-22.56\%} & \multicolumn{2}{c|}{0.0049} & \multicolumn{2}{c|}{-54.90\%} & \multicolumn{2}{c|} {0.0114} & \multicolumn{2}{c|}{-40.30\%} & \multicolumn{2}{c|} {0.0085}& \multicolumn{2}{c|}{-48.98\%} & \multicolumn{2}{c|} {0.0099} & \multicolumn{2}{c|}{\textbf{-58.87\%}} & \multicolumn{2}{c} {\textbf{0.0133}} \\
    \hline
     
    {ClassC} &  \multicolumn{2}{c|}{-19.57\%} & \multicolumn{2}{c|}{0.0054} & \multicolumn{2}{c|}{-24.89\%} & \multicolumn{2}{c|}{0.0081} & \multicolumn{2}{c|}{-43.11\%} & \multicolumn{2}{c|} {0.0133} & \multicolumn{2}{c|}{-35.51\%} & \multicolumn{2}{c|} {0.0133}& \multicolumn{2}{c|}{-42.99\%} & \multicolumn{2}{c|} {0.0157} & \multicolumn{2}{c|}{\textbf{-53.08\%}} & \multicolumn{2}{c} {\textbf{0.0186}} \\
    \hline
     
    {ClassD} &  \multicolumn{2}{c|}{-9.68\%} & \multicolumn{2}{c|}{0.0023} & \multicolumn{2}{c|}{-22.44\%} & \multicolumn{2}{c|}{0.0067} & \multicolumn{2}{c|}{-43.64\%} & \multicolumn{2}{c|} {0.0123} & \multicolumn{2}{c|}{-32.33\%} & \multicolumn{2}{c|} {0.0093}& \multicolumn{2}{c|}{-43.55\%} & \multicolumn{2}{c|} {0.0151} & \multicolumn{2}{c|}{\textbf{-55.99\%}} & \multicolumn{2}{c} {\textbf{0.0182}} \\
    \hline
     
    {ClassE} &  \multicolumn{2}{c|}{-30.82\%} & \multicolumn{2}{c|}{0.0018} & \multicolumn{2}{c|}{-29.08\%} & \multicolumn{2}{c|}{0.0027} & \multicolumn{2}{c|}{-58.63\%} & \multicolumn{2}{c|} {0.0048} & \multicolumn{2}{c|}{-41.30\%} & \multicolumn{2}{c|} {0.0041}& \multicolumn{2}{c|}{-51.11\%} & \multicolumn{2}{c|} {0.0046} & \multicolumn{2}{c|}{\textbf{-65.75\%}} & \multicolumn{2}{c} {\textbf{0.0056}} \\
    \hline
     
    {UVG} &  \multicolumn{2}{c|}{-37.50\%} & \multicolumn{2}{c|}{0.0056} & \multicolumn{2}{c|}{-16.46\%} & \multicolumn{2}{c|}{0.0046} & \multicolumn{2}{c|}{-53.87\%} & \multicolumn{2}{c|} {0.0100} & \multicolumn{2}{c|}{-37.52\%} & \multicolumn{2}{c|} {0.0068}& \multicolumn{2}{c|}{-43.29\%} & \multicolumn{2}{c|} {0.0078} & \multicolumn{2}{c|}{\textbf{-59.89\%}} & \multicolumn{2}{c} {\textbf{0.0121}} \\

    \hline
    %\bottomrule[2pt]
  \end{tabular}
  \label{tab:BDrate_ssim}
\end{table*}

\subsection{Results}
Figure~\ref{fig:different_decoders} demonstrates the intermediate predictions by the independent decoders in our HMC approach. As can be observed, in regions containing irregular motions and occlusions, adaptive kernel-based prediction yields more plausible prediction (albeit blurred) than the vector-based prediction. This is because the kernel-based motion decoder generates multiple offsets and attention maps for different groups of feature maps, and can handle these challenging areas better. On the other hand, in areas with slow and simple motion or stationary regions, the vector-based prediction is sharper and more accurate. As also demonstrated in Fig.~\ref{fig:one-to-one prediction}, adaptive kernel-based prediction generally provides significant improvement over the vector-based prediction, and the weighted combination of these two predictions yields additional gain over using only adaptive kernel-based prediction.

\textbf{Performance in terms of PSNR.} As shown in Fig.~\ref{fig:rd_psnr}, our PSNR model outperforms all the existing learning methods in the entire bit rate range. At high bit rate, our method achieved a significant gain, over 1dB compared with HEVC. In terms of BD-rate, our model achieved 55.61\% bit rate reduction against the H.264 anchor, surppassing  the methods of Lu {\it et al.}~\cite{lu2020end} (52.56\%), Agustsson {\it et al.}~\cite{agustsson2020scale} (45.78\%) and Yang {\it et al.}~\cite{Yang_2020_CVPR} (42.98\%) on UVG. More results are shown in Tab.~\ref{tab:BDrate}.

\textbf{Performance in terms of MS-SSIM.} As shown in Fig.~\ref{fig:rd_ssim}, our MS-SSIM optimized model achieves the state-of-the-art performance and shows consistent performance gain across all the video categories. Using H.264 as an anchor, our MS-SSIM model have 59.89\% bit rate saving in comparison with Agustsson {\it et al.}~\cite{agustsson2020scale} (53.11\%), Liu {\it et al.}~\cite{liu2019learned} (49.73\%), Yang {\it et al.}~\cite{Yang_2020_CVPR} (46.18\%) and Lu {\it et al.}~\cite{lu2020end} (43.29\%) on UVG. Table~\ref{tab:BDrate_ssim} releases the remaining results and shows the consistent gains across all the testing datasets.

\begin{figure}[t]
\centering
\includegraphics[scale=0.32]{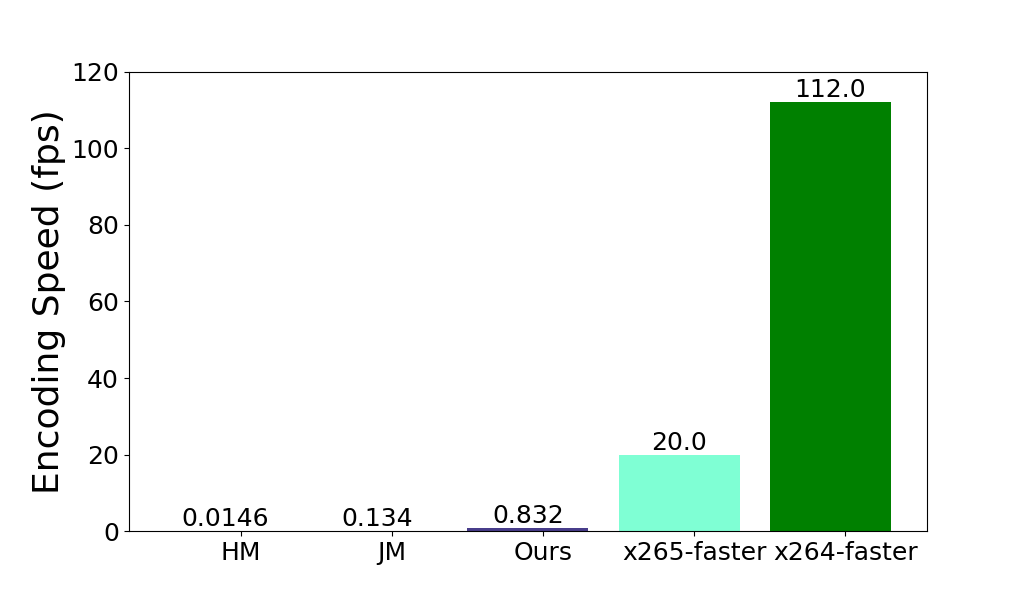}
\caption{{Encoding Speed Analysis.}}

\label{coding_speed}
\end{figure}

\textbf{Computation Time Analysis.} We have evaluated the coding time of different methods for comparison as shown in Fig.~\ref{coding_speed}.  We conduct several experiments on  the  Intel  Xeon  E5-2680  v4  CPU@2.40GHz  and  a  TiTan XP GPU. We select two common official softwares JM and HM, two commercial softwares x264 and x265. Our proposed method provides 0.832fps for encoding a 1080p video which is nearly 50 times faster than HM. The x264and x265 are mainly optimized for practical and commercial use. Network acceleration and simplification can also be applied for fast and real-time applications.
% 6-27 3:10

\subsection{Ablation Study}

\begin{figure}[t]
\centering
\includegraphics[scale=0.285]{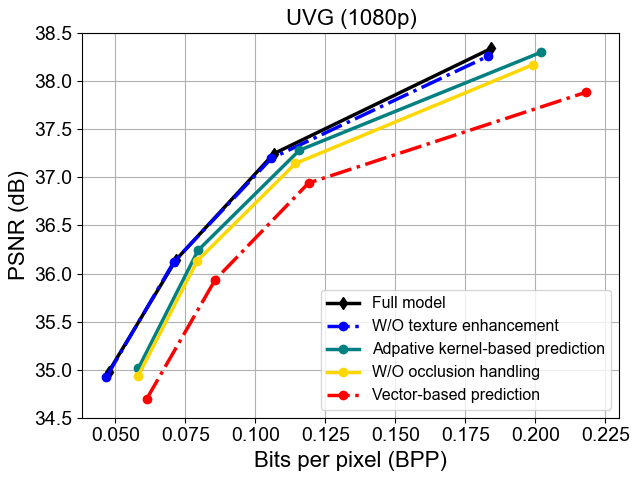}
\caption{{Performance comparison with different motion compensation approaches.}}
\label{fig:one-to-one prediction}
\end{figure}

\textbf{Benefits of hybrid motion compensation.} To evaluate the relative performance of different motion compensation strategies, we have trained three alternative models: 
1) hybrid prediction without using the texture enhancement component, 2) using only vector-based prediction, and 3) using only adaptive kernel-based prediction. Figure~\ref{fig:one-to-one prediction} compares the performance of different models on the UVG dataset. Removing the texture enhancement component will have marginal degradation.  Using adaptive kernel-based prediction only will reduce the performance slightly by 0.2dB, whereas using vector-based prediction only will lead to significant performance drop about 0.5dB.

\begin{figure}[t]
\centering
\includegraphics[scale=0.285]{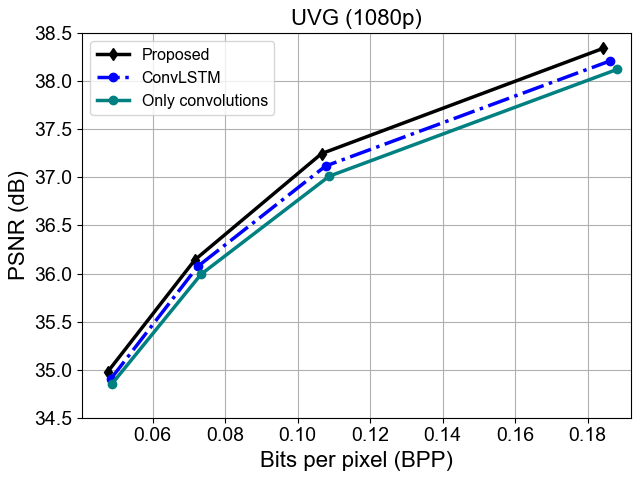}
\caption{{Comparison of different approaches for information aggregation to generate the compound motion representation.}}
\label{fig:temporal_efficiency}
\end{figure}
\textbf{Impact of recurrent information aggregation.} To verify the efficiency of the residual ST-LSTM for generating the CSTR, we also trained models where the RIA module is replaced with a conventional ConvLSTM and stacked convolution layers (no temporal recursion or convolution, and using only spatial convolution for the extracted features from the encoder), respectively. As shown in Fig.~\ref{fig:temporal_efficiency}, without the spatiotemporal memory  (i.e., using ConvLSTM), the quality of the reconstructed frame is reduced on average by 0.1dB at high bit rate. The performance will degrade about 0.1dB further when we just apply multiple layers of spatial convolutions to quantized spatiotemporal feature, which is generated from the current frame and one previous decoded frame. When the bit rate is lower, the benefit from using multiple frames is reduced because the spatiotemporal feature buffer contains less information at the current bit rate and does not affect the CSTR generation compared with the high bit rate fusion.

\textbf{Occlusion aware activation for feature alignment.} In the adaptive kernel-based prediction, to better handle the occlusions, we decode attention maps following the approach of~\cite{Gui_2020_CVPR} in addition to offset maps to enable implicit feature activation for different regions. Without the attention maps for implicit occlusion masking, the PSNR will drop about 0.12dB as shown in Fig.~\ref{fig:one-to-one prediction}.

%\textbf{Random noise regularization for training.} We also removed RNR to train our model and found a obvious performance drop in the entire bit rate range, as shown in Fig.~\ref{fig:one-to-one prediction}. It is mainly caused by the inconsistency between training and testing with the change of quantization.

\begin{figure}[t]
\centering
\includegraphics[scale=0.145]{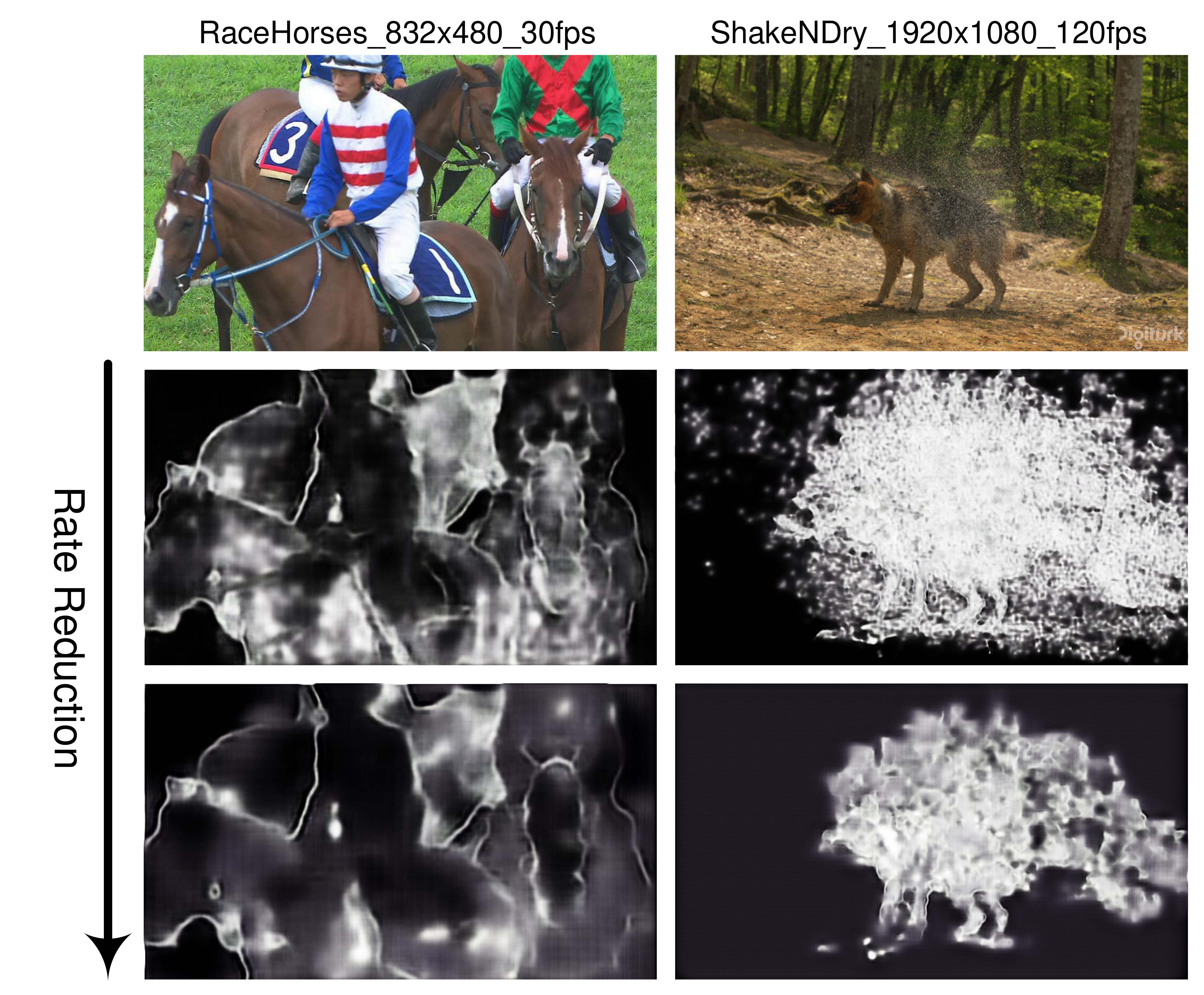}
\caption{Effects of the bit rate on the motion compensation mode. While moving edges are predicted using the adaptive kernel-based resampling over the entire rate regime, the interior of moving regions switch to using optical flow based  resampling at the lower bit rate.}
\label{fig:weights_selection}
\end{figure}
\textbf{Low and High bit rate compensation mode selection.} Recall that all the decoded information for motion compensation is constrained by the bit allocation for the compound representations. We have analyzed how does the motion compensation mode decoder adaptively switch the mode in different regions at different bit rates in Fig.~\ref{fig:weights_selection}. Over the entire rate range, the adaptive kernel-based prediction can better handle occlusions around motion edge and regions with fast motion as shown in white regions. As the bit rate is reduced, the regions predicted mainly by kernel-based motion decoder will gradually shrink but the kernel-based prediction is still favored over moving edges. We suspect that this is because the decoded reference frame is already blurred at the low bit rate, so that the vector-based prediction suffers from  less artifacts than at the higher rate.

\section{Conclusion}\label{sec:conc}

In this paper, we proposed a novel inter-frame coding scheme that utilizes a rate-constrained compound spatiotemporal representation to perform hybrid motion compensation (HMC). Multiple predictions are generated through a one-to-many decoding pipeline including vector-based predicted frame, kernel-based predicted frame, mode selection maps and texture enhancement. The HMC module adaptively combines the predictions obtained using pixel-level alignment based on vector-based resampling and the prediction acquired through feature-level alignment based on the adaptive kernel-based resampling. The former prediction is more accurate in regions with small and regular motions, while the latter prediction is more accurate in the occluded regions near the moving edges.

The proposed inter-coding scheme is integrated in an end-to-end video coding framework. Our extensive simulation results show that the proposed framework is either on-par or outperforms all the existing learning-based approaches as well as the traditional video codecs H.264 and H.265, in terms of both PSNR and MS-SSIM, over a large rate range.

Although the proposed inter-coder has lead to promising results, more ablation studies are needed to understand the trade-offs between the performance and the complexity under different settings to yield a computationally efficient model. The texture enhancement component does not appear to bring significant gains and may be skipped for reduced complexity. To better solve the problems of performance consistency among different datasets with different resolutions and frame rate, we can further apply multi-scale representations in HMC.

 The proposed coder achieved large performance gain at the high rate regime, but the performance is similar to  the H.265 at the low rates, as with other leading learning-based coders. This is likely because the learnt inter-coder suffers from error propagation in the reconstruction of successive P-frames. One important future direction in learning-based video coding is to design effective error propagation minimization strategies such as multiple model adaptation.

\section*{Acknowledgement}

\bibliographystyle{IEEEtran}
\bibliography{nvc}

% \ifCLASSOPTIONcaptionsoff
%   \newpage
% \fi

\end{document}